\documentclass[twocolumn]{aastex6}
\pdfoutput=1 
\usepackage{amsmath,amstext}
\usepackage[T1]{fontenc}
\usepackage{apjfonts} 

\shorttitle{Fallback SN in binaries}
\shortauthors{Schr\o der {\em et al.}}

\begin{document}

\title{Black Hole Formation in Fallback Supernova and the Spins of LIGO Sources}

\author{ Sophie L. Schr\o der, Aldo Batta and
Enrico Ramirez-Ruiz}

\affil{Niels Bohr Institute, University of Copenhagen, Blegdamsvej 17, 2100 Copenhagen, Denmark \\ Department of Astronomy and Astrophysics, University of California, Santa Cruz, CA 95064, USA}

\begin{abstract}
Here we investigate within the context of field binary progenitors how the the spin of LIGO sources vary when the helium star-descendent black hole (BH) is formed in  a failed supernova (SN) explosion rather than by direct collapse.  To this end, we make use of 3d hydrodynamical simulations of fallback supernova in close binary systems with properties designed to emulate LIGO sources. By systematically varying the explosion energy and the binary properties, we are able to explore the effects that the companion has on redistributing the angular momentum of the system.  We find that, unlike the mass,  the spin of the newly formed BH varies only slightly with the currently theoretically unconstrained energy of the SN  and is primarily  determined by the initial binary separation.
In contrast,  variations in the initial  binary separation yield sizable changes on the resultant effective spin of the system. This implies that the formation pathways of LIGO sources leading to a particular effective spin might be far less restrictive than the standard direct collapse scenario suggests.
\keywords{close binaries, supernova, GW sources}
\end{abstract}


\section{INTRODUCTION}
\label{intro}
The gravitational wave (GW) signals detected  by LIGO \citep{LIGO_GW151226,LIGO_GW150914, LIGO_GW170104,LIGO_GW170608,LIGO_GW170814}  have uncovered a population of black holes  (BHs) that is significantly  more massive than the population known to reside in accreting binaries \citep{Remillard2006}.  While there is significant debate in the community about how black hole binaries are assembled 
\citep{Zwart_2000, Kalogera_2007, Sadowski_2008, Postnov_2014, GW150914_Implications, Belczynski_2016, Rodriguez_2016b, deMink_2016, Gerosa2017, Wysocki2017}, the {\it classical scenario} \citep{Tutukov_1993,Voss_2003} remains one of the leading candidates. In this channel, a wide massive binary undergoes a series of mass transfer episodes leading to a tight binary comprised of a massive helium star ($M_{\ast}$) and a BH ($M_1$), prior to the formation of the second BH ($M_2$). LIGO observations of the mass-weighted angular momentum perpendicular to the orbital plane $\chi_{\rm eff}$, have been argued to provide constraints on this formation channel \citep{Rodriguez_2016c,Farr_2017,Stevenson2017}. This is because vital information on the mass transfer history of the binary and the spin of $M_\ast$ is imprinted on  
\begin{equation}
\chi_{\rm eff}={ M_1 \vec{a_1}+ M_2 \vec{a_2}  \over M_1 + M_2 } \cdot \hat{L}.
\end{equation}
Here $\vec{a_1}$ and $\vec{a_2}$ are the dimensionless spins of the BHs and $\hat{L}$ is the direction of the angular momentum in the orbital plane. 

The angular momentum of  the secondary  BH is intimately linked to that of the  progenitor helium star, which in turn is determined by its mass-loss history  and the torque exerted by the primary black hole \citep{Qin_2018,Kushnir_2016,Zaldarriaga_2018}. This torque can effectively drive synchronization of the stellar spin and the orbit in binaries tighter than $d_{\rm \tau }$, the maximum separation allowed for synchronization within the life of the helium star \citep{Zaldarriaga_2018}. The final angular momentum of the star thus provides a reasonable estimate of $a_2$ when the mass and angular momentum losses from the final supernova explosion are ignored.  All previous works  have assumed  that such effects are small based on the simple expectation that LIGO BHs are formed by direct collapse.  

Motivated by the fact that the nature of BH-forming supernova  (SN) explosions is not to well understood \citep{Fryer2012,Ugliano2012,Pejcha2015a}, in this {\it Letter} we explore the effects  on $\chi_{\rm eff}$ when the second BH $M_2$ instead is formed by a fallback SN explosion  \citep{Moriya2010,Fryer2012, Dexter2013,Lovegrove2013,Perna2014,Batta_2017,Fernandez_2018}.  For this purpose we make use of 3d hydrodynamical simulations of fallback  SN in  close binary systems with properties aimed at reproducing LIGO GW signals.  The structure of this {\it Letter} is as follows. In Section \ref{section:Method} we describe the numerical formalism used to initiate the fallback SN explosion and compute  the subsequent evolution of the binary.  In Section  \ref{section:Results}  we describe the  dynamics of the fallback material and its effect on the final spin of both BHs.  Lastly, in Section \ref{section:discussion} we present our key findings and relate them to the current population of LIGO sources.

\section{METHODS AND INITIAL SETUP}
\label{section:Method}

\begin{figure}
\begin{center}
  \includegraphics[trim={1.0cm 0.2cm 0.6cm 0.8cm},width=1.05\linewidth]{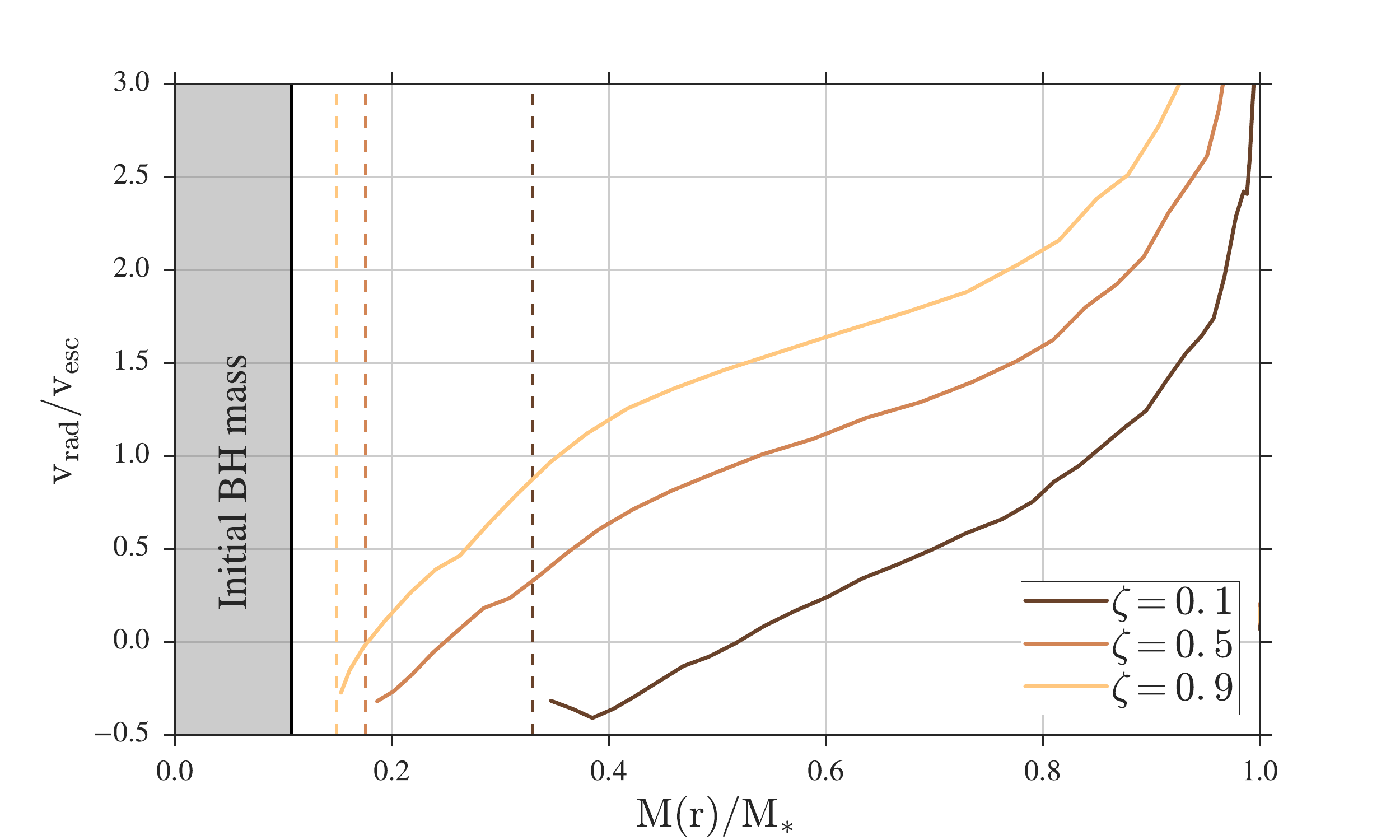}
\caption{The velocity profile of the ejecta at $d_\ast=1$ (soon after  the shock emerges from the stellar envelope) is plotted  in units of the escape velocity for three different initial explosion energies.The shaded region shows  the initial BH mass. The subsequent growth of the  BH depends on the amount of fallback material, which in turn, depends strongly on the energy injected.  The {\it dashed} lines show  the current mass of the BH for different explosion energies.}
\label{fig:velprofiles}
\end{center}
\end{figure}

Here we follow the setup described in \citet{Batta_2017} to study the evolution of the progenitor binary system after the  birth of $M_2$.  We make use of a modified version of the three-dimensional  smoothed particle hydrodynamics (SPH) code {\tt GADGET2} \citep{Springel_2005}, with our initial setup consisting of  a tidally locked $28 M_{\odot}$ helium star in orbit around a  BH of $M_1=15 M_{\odot}$. The reader is referred to \citet{Batta_2017} for further details on initial particle distribution and numerical accretion. We settled for a resolution of  $5\times 10^{5}$ particles, which showed convergence for the accretion rates and properly captured the dynamics of the ejecta and the binary system.

To study the interaction of the fallback material with the newly formed BH binary we explore three sets of simulations. Each set starts with the initial binary in a circular orbit with a separation $d = d_\ast R_{\ast}$, where  $d_\ast = 2$, $3$ or  $5$ and $R_\ast$ is the pre-SN star's radius. Then for each orbital separation we run simulations with at least four different SN explosion energies.  In all cases we assume $a_1=0$ based on  the results of \cite{Qin_2018}, and assume that synchronization of the stellar spin and the orbit has taken place, as is expected for the initial separations used in this analysis. Given the large uncertainties in BH natal kick estimations \citep{Mandel_2016b, Repetto_2015}, we assumed the simplest scenario where no natal kick is applied to the recently formed BH. This combined with the synchronization of the stellar spin and the orbit, translates into BH's spins aligned with the orbital angular momentum.

The initial profile of the star was obtained from the 35OC {\tt KEPLER} model calculated by  \citet{Woosley_2006}  of a $28 M_{\odot}$ pre-SN helium star with  $R_\ast=0.76 R_\odot$.  We considered the innermost $3M_\odot$ of the pre-SN star to be the newly formed BH with  $a_2 (t=0) = 0$, which we subsequently  treat as a sink particle. After the removal of the  inner core, we use a parameterized energy injection routine to mimic the supernova engine  and derive the density and velocity profile of the expanding envelope. Specifically, we use a spherically symmetric kinetic piston at the inner boundary to adjust the energy injected into the envelope. In all calculations the  energy is parametrized as follows: $\mathrm{E_{\rm SN} }= \zeta \ \mathrm{E_{\rm G} }$, where  $E_{\rm G}=2.3 \times 10^{52} {\rm erg}$ is the binding energy of the pre-SN star.  

The distribution that describe the ejecta is determined solely by the structure of the pre-SN star \citep{Woosley_1995, Matzner_1999} and is established by the hydrodynamics of the interaction. Initially, the shock propagates through the stellar material, pressurizing it and setting it into motion. Once the shock wave approaches the surface of the star, a rarefaction  stage begins  in which stellar material is accelerated by  the entropy deposited by the shock. This stage terminates once the pressure ceases  to be dynamically important and the material expands  freely.   Figure \ref{fig:velprofiles} shows the radial velocity profile of the envelope when the shock surfaces  the stellar envelope for three different explosion energies: $\zeta=0.1, 0.5$ and 0.9.  The gray area shows the initial BH mass  while the dashed lines show $M_2(t)$ at the time the shock reaches $r =R_\ast$. Despite the complicated hydrodynamical interaction, the density and pressure approach steep power laws in velocity as the material approaches homologous expansion (Figure \ref{fig:velprofiles}).  For a given progenitor structure, the ensuing ejecta will take similar distributions. 

\begin{figure*}[t]
  \centering
    \includegraphics[trim={0.0cm 1.0cm 0.0cm 0.0cm},clip,width=1.0\linewidth]{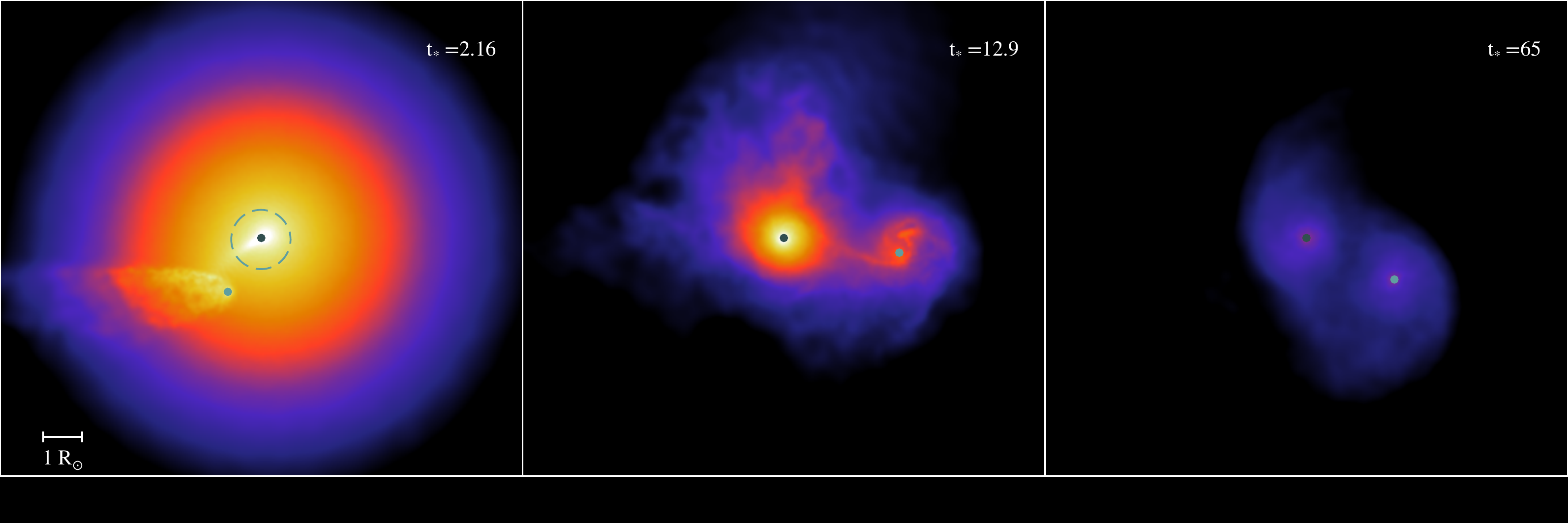}
    
    \includegraphics[trim={0.0cm 1.0cm 0.0cm 0.0cm},clip,width=1.0\linewidth]{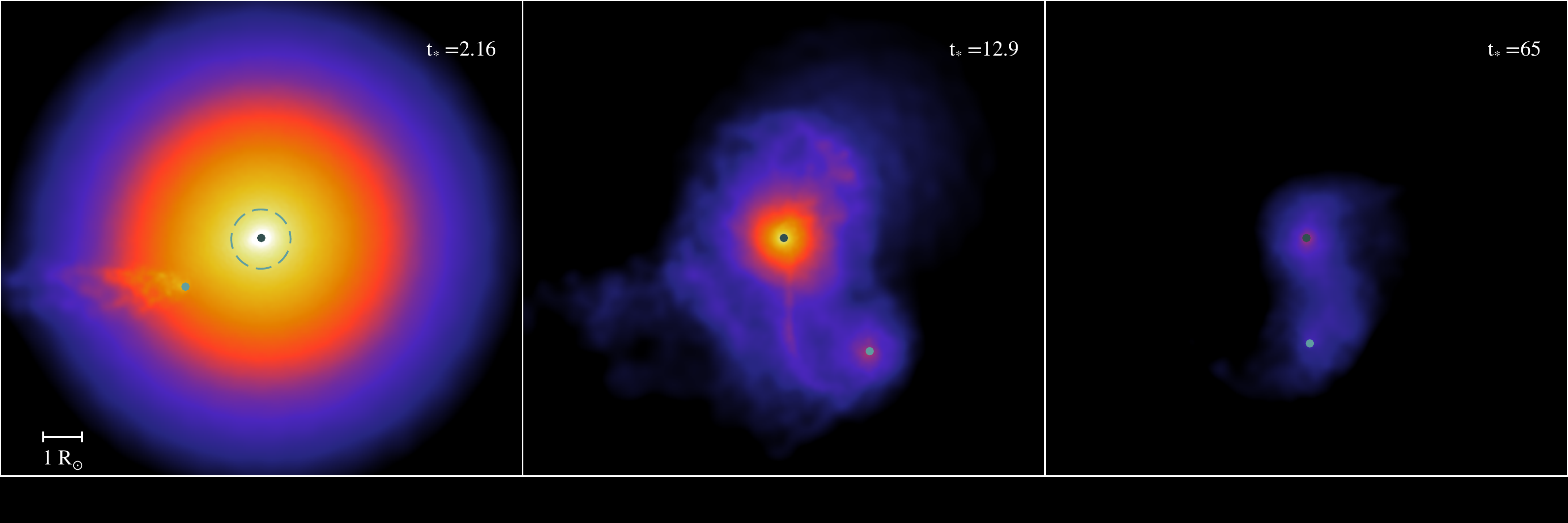}  
    
    \includegraphics[trim={0.0cm 0.0cm 0.0cm 0.0cm},clip,width=1.0\linewidth]{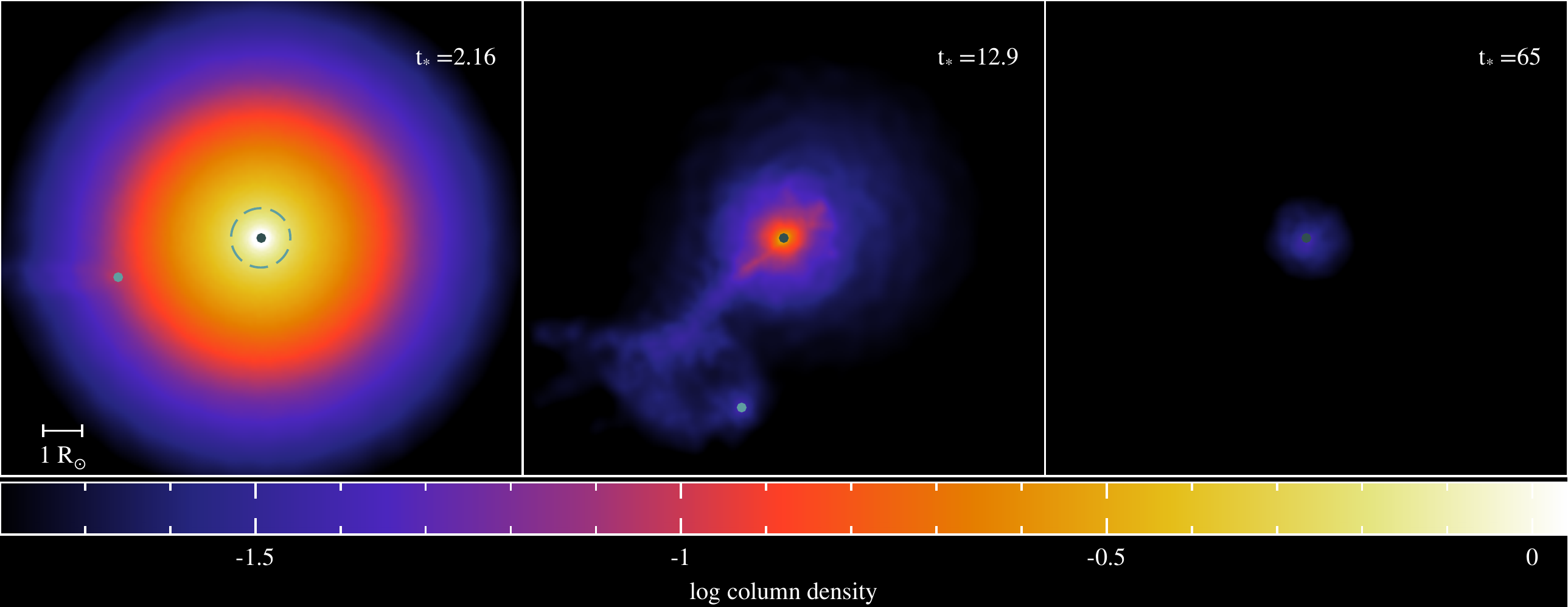}  
  \caption{Gas column density (code units) in the equatorial plane of the binary for three different initial separations  (from {\it top} to {\it bottom}) and at three different evolutionary stages ({\it left} to {\it right}).The frames are centered on the newly formed BH ({\it black} circle) and the dashed circles in the left panels show the initial size of the pre-SN star. The {\it cyan} circle  shows  the companion BH. Times are measured in units of $t_\ast$. All simulations have $\zeta  = 0.4$ but differ on the initial separation of the binary: $d_\ast = 2$ ({\it top} panel), $d_\ast = 3$ ({\it middle} panel) and $d_\ast = 5$ ({\it bottom} panel). In the last frame in the {\it bottom} panel, the resulting large binary separation places  the companion outside the frame. }
  \label{fig:separations}
  \vspace*{0.4cm}
\end{figure*}

 When $\zeta \lesssim 1$, energy injection fails to  unbind the star such that a sizable fraction of its mass eventually  fallbacks onto the newly formed BH. If this takes place in a binary system \citep{Batta_2017}, a non-negligible  fraction of the bound  material can expand to a radius comparable or larger than the binary's separation, thus immersing the BH companion in gas. The interaction of fallback material with the binary transfers orbital angular momentum to the gas, which upon accretion onto the orbiting BHs is  finally transferred into spin angular momentum. Differences in $\zeta$ result in diverse accretion histories, which ultimately regulate the BHs' final masses and spins. It is to this topic that we now draw our attention.
 
\section{Fallback Supernova in Binaries and the Spins of LIGO Sources}
\label{section:Results}

\subsection{Spin Evolution of the Newly Formed Black Hole}

\begin{figure*}
\makebox[\textwidth][c]{
  \includegraphics[trim={1.6cm 0.2cm 2.8cm 1.2cm},clip,width=1\linewidth]{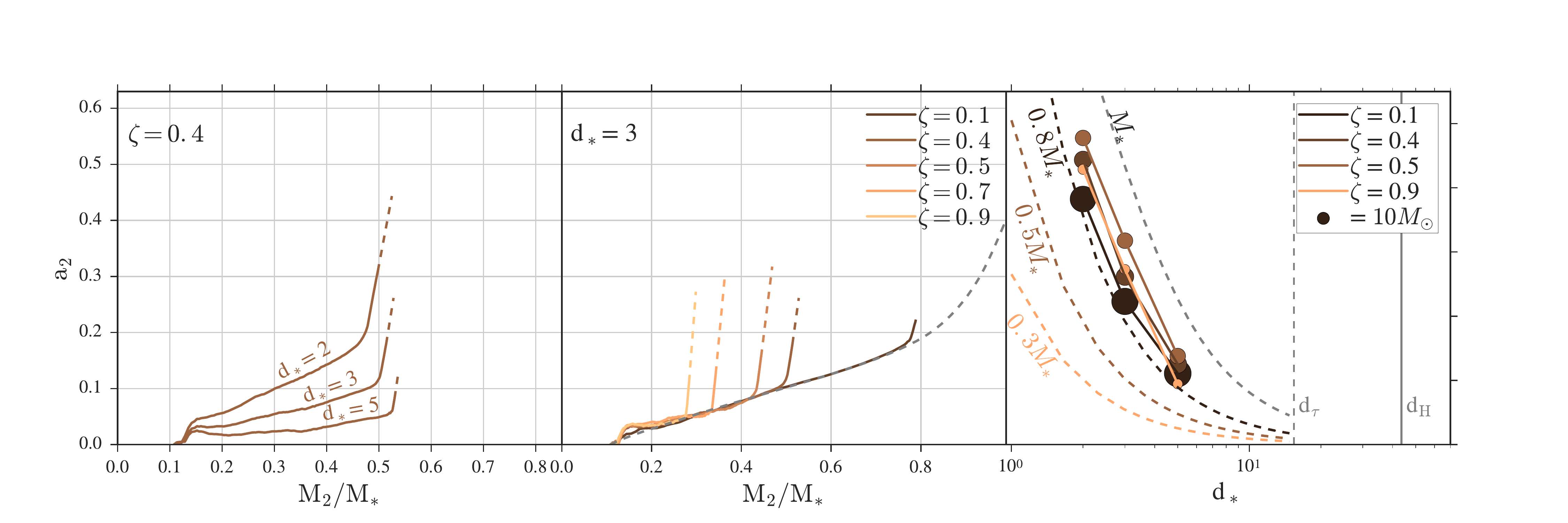}}
  \vspace*{-0.4cm}
\caption{The dependence of $a_2$ on $\zeta$ and $d$. {\it Left panel:} The spin parameter of $M_2$ as a function of the accreted mass in units of $M_{\ast}$ for the three simulations shown in  Figure \ref{fig:separations}. Here all simulations have  $\zeta = 0.4$ and, as a result, the total accreted mass is similar. The {\it solid} lines correspond to the values derived from the simulations while the {\it dashed} lines indicates the expected mass and angular momentum accretion from material that remains in the disk \citep{Bardeen_1970,Thorne_1974} when  the  simulation ends at $t = 80 t_{ \ast}$.
{\it Middle panel:} The dependence of $a_2$ on $\zeta$. All simulations shown have  the same separation ($d=3R_\ast$) but $\zeta = 0.1$, $0.4$, $0.5$, $0.7$ and $0.9$. The {\it dashed} line represents the spin expected from direct collapse (i.e., $\zeta=0$). {\it Right panel:} The dependence of $a_2$ on $d_\ast$ and $\zeta$. The size of the symbols show the final BH mass, ranging from $8.2 M_\odot$ to $22.5 M_{\odot}$. The vertical  {\it dashed} line shows $d_\tau$, the radius for effective tidal synchronization \citep{Zaldarriaga_2018} and the  vertical {\it solid} line shows $d_{\rm H}$, the binary  separation required for merging within  a Hubble time ($28 M_\odot$ + $15 M_\odot$). {\it Dashed} lines depict the spin expected from the direct collapse of different mass fractions,  $f M_\ast$, of the  pre-SN star with $f=0.3,0.5,0.8, 1$.}
\label{fig:Mbhspin}
\end{figure*}

Figure \ref{fig:separations} shows the gas column density in the equatorial plane of the binary for three different simulations (from {\it top} to {\it bottom}) and  at three different stages ({\it left} to {\it right}). Evolutionary times  in Figure \ref{fig:separations} are measured in units of the dynamical time of the pre-SN star: $t_\ast$. All simulations have $\zeta = 0.4$  but differ on the initial separation of the binary: $d_\ast = 2$ ({\it top} panel), $d_\ast=3$ ({\it middle} panel) and $d_\ast=5$ ({\it bottom panel}). The frames are centered on the newly formed BH and the dashed circles in the {\it left} panels show the  size of the pre-SN star.

Flow dynamics are similar in all three simulations shown  in Figure \ref{fig:separations}. First, the envelope expands to rapidly engulf  the companion BH.  A bow shock is created as a result of this initial  interaction. It is, however, only when the slower moving material  reaches the companion  that the resulting  torque can  supply the envelope gas with sizable  angular momentum. This envelope material will remain bound to the system and will  form a disk around  $M_2$ if restricted  to  the region within which orbiting gas is gravitationally bound to the newly formed BH.  A  disk, albeit lighter,   also forms around $M_1$, whose final mass depends sensitively on the initial separation.

The total mass bound to  $M_2$  is the same  in all simulations, yet the fraction of  angular momentum accreted  increases with decreasing separation.  As a result, $a_2$ is higher  for progressively  more compact binaries despite the final mass of $M_2$  reaching similar values. This can be  seen in the {\it left} panel of Figure \ref{fig:Mbhspin}, in which  we show the evolution of $a_2$  as a function of the accreted mass in units of $M_{\ast}$. Initially, $a_2$ increases as envelope material  is accreted directly  onto the BH. The  innate angular momentum in this initial phase is determined by tidal synchronization, which increases as the binary separation decreases \citep{Kushnir_2016,Kushnir_2017,Zaldarriaga_2018}. A transition in the evolution of $a_2$ is observed in the {\it left} panel in Figure \ref{fig:Mbhspin} when material that is effectively torqued by the binary is able to form a disk and  is subsequently accreted onto $M_2$.   This material has a higher specific angular momentum than the one initially set by tidal synchronization  and, when accreted, is able to spin up the newly formed BH at a faster rate.  The resultant  change in slope  observed  in the {\it left} panel in Figure \ref{fig:Mbhspin}  due to the accretion of disk material is observed to occur earlier for  smaller separations, which results in higher total spins values than those given by direct collapse of the same fallback material.

At a fixed  $\zeta$, the spin of the newly formed BH depends sensitively  on $d_\ast$. For $d_\ast \lesssim 5$, the resultant torque on the fallback material can be considerable and, as a result, $a_2$ can be appreciable  larger  than the one expected from tidal synchronization.  In this case, the final mass of $M_2$ remains unchanged while the final spin can vary drastically.  The final mass of $M_2$ is, on the other hand, controlled by  $\zeta$.  The {\it middle} panel  in Figure \ref{fig:Mbhspin} shows the evolution of $a_2$ for a fixed separation $d_\ast=3$  and changing $\zeta$.  Initially, the spin evolution follows the trend expected from direct collapse. This is because  the torque  is unable to  modify the original  angular momentum of the promptly collapsing stellar material.  A transition to disk accretion is seen in all cases, with the shift always occurring  late in the mass accretion  history of $M_2$.  The resultant  spin  is similar in all cases due to the self similarity of the mass distribution of the expanding ejecta \citep{Matzner_1999}, which results in a comparable mass ratio of directly falling stellar material to disk material for different values of $\zeta$. 
For example this fraction varies from 5.3 \% for $\zeta = 0.5$ to 7.6\% for $\zeta = 0.9$ (see $middle$ panel of Figure \ref{fig:Mbhspin} for $d_\ast = 3$). 
This mass ratio is mainly responsible for determining  the final spin of the BH and varies only slightly with $\zeta$.

We  have discussed, in the context of the {\it classical scenario},  the effects that the binary separation and the energy of the SN have on the resulting spin of the newly formed BH. The {\it right} panel of Figure \ref{fig:Mbhspin} provides a clear summary of our findings as it shows the  final spin of $M_2$ as function of $d$ and $\zeta$.  The final mass of the newly formed  BH is also shown by  the size of the symbols. The masses for $M_2$  range from $22.5 M_{\odot}$ for $\zeta=0.1$  ($M_2/M_\ast \approx 0.8$) to $8.2 M_\odot$ for $\zeta=0.9$  ($M_2/M_\ast \approx 0.3$).  Together with the results from our simulations we also plot the expected spin obtained from the direct collapse of the pre-SN stellar profile. This formalism makes use of the {\tt KEPLER} model and assumes solid body rotation determined by tidal synchronization.  Then, by assuming the spherical collapse of the star, we obtain the BH's spin $a_2$ for different fractions $f=0.3,0.5,0.8,1$ of the collapsed stellar mass $M_\ast$. If the entire star was to collapse directly onto a BH,  this will give a final spin $a_2$ solely dependent on $d$, as predicted by the {\it dashed} line in Figure \ref{fig:Mbhspin} labeled  $M_\ast$.

When $\zeta$ is small and a significant fraction of the material is promptly accreted by the BH, the simple direct collapse formalism provides an accurate description of the final spin of the newly formed BH. This can be seen by comparing the  {\it dashed} line  in Figure \ref{fig:Mbhspin} labeled  $0.8 M_\ast$ with the simulation results obtained for  $\zeta=0.1$, which give  BHs with   $M_2\approx 0.8 M_\ast$ and final spins that closely resemble the direct collapse ones. By contrast, when $fM_\ast  \lesssim M_\ast$, the final spin is significantly higher than the one predicted by direct collapse of the same enclosed material. This is because in such cases the fallback material is effectively torqued by the BH companion, which results in disk formation and consequentially higher final spin values. Binary BH formation in the {\it classical scenario}  depends critically on the  currently poorly constrained energy of the resulting SN, which for fallback-mediated remnant growth results  in  faster spinning BHs than what would have been attainable for a single star progenitor.

\subsection{Spin Evolution of the Orbiting Black Hole}
\begin{figure}[t]
  \makebox[\textwidth][l]{
  \includegraphics[trim={0.8cm 0.8cm 1.0cm 3.0cm},clip,width=0.9\linewidth]{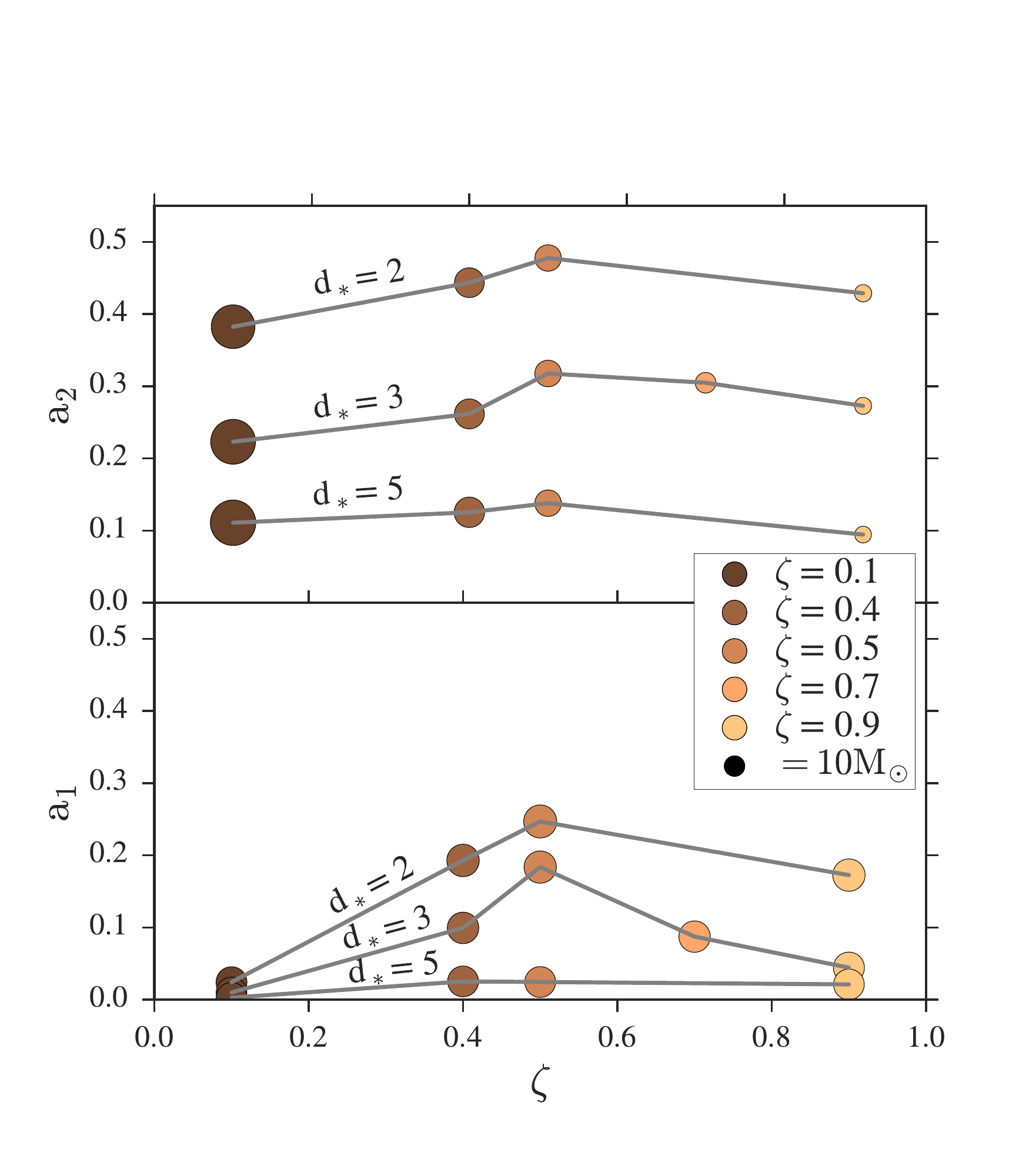}}
\caption{The dependence of $a_1$ and $a_2$ on both $\zeta$ and $d$. The size of the symbols depicts  the value of  the final mass of the BH.
 {\it Top panel}: $a_2$ as a function  $\zeta$ for initial  $d_\ast = 2$, $d_\ast = 3$ and $d_\ast = 5$. 
{\it Bottom panel}:  $a_1$ as a function  $\zeta$ for initial  $d_\ast = 2$, $d_\ast = 3$ and $d_\ast = 5$, under the  assumption that  $M_1$ had  no spin before the final SN explosion $a_1 (t = 0) = 0$.}
\label{fig:comparison}
\end{figure}

Figure \ref{fig:comparison} shows the dependence of $a_1$ and $M_1$ on  $\zeta$ and $d$. In contrast to $M_2$,   the final mass of the companion BH is only weakly altered by changes in $\zeta$. The reason is that  a comparatively small mass can be effectively  restricted  to  the region within which the expanding envelope material  is gravitationally bound to $M_1$.  This bound material forms a disk whose final mass depends on  both $d$ and $\zeta$.  The resultant  changes in $a_1$, under the assumption of $a_1(t=0)=0$, are  observed  to  be more pronounced when the initial binary  separation changes. Although, as expected, no sizable changes  take place at large separations given that  only a tiny fraction of the companion's  envelope can be under the gravitational influence of $M_1$. The final value of $a_1$ shows a modest  variation with SN energy with  a small preference for $\zeta \approx  0.5$  at small separations. This indicates that although the ejecta distributions are similar for changing values of $\zeta$, the fraction of bound material to $M_1$ is largest for this particular explosion energy, although its  exact value  is likely to change  for different pre-SN progenitors. 

\section{DISCUSSION}
\label{section:discussion}
In this {\it Letter} we have explored  within the {\it classical}  binary scenario how the spin of LIGO sources vary when  the remnant BH is formed in weak   SN explosions instead of direct collapse.  Our key findings are summarized below.

\begin{itemize}

\item The final mass of the newly formed BH depends on the explosion energy. Its  mass varies from $M_2 \approx 0.8M_\ast $ for $\zeta=0.1$  to $M_2 \approx 0.3M_\ast $ for $\zeta=0.9$ (Figure~\ref{fig:velprofiles}). 

\item At a fixed SN energy, the final spin increases significantly  with decreasing $d$ as a larger fraction of  the fallback material is torqued by the companion. This results in similar mass BHs but with widely different spins (see {\it left} panel in Figure~\ref{fig:Mbhspin}).

\item Due to the self similarity of the mass distribution of the expanding ejecta, the final spin of the BH varies only slightly with $\zeta$. This results in BHs with a wide range in masses but similar spins (see {\it middle} panel in Figure~\ref{fig:Mbhspin}).

\item In the presence of a companion,  the final spin of a BH formed by a  fallback SN explosion can be significantly higher than the one predicted by direct collapse of the same stellar  material (see {\it right} panel in Figure~\ref{fig:Mbhspin}).  

\item The spin of the BH companion, on the other hand,  depends on both  $\zeta$ and $d$. This is because its  accretion history  is determined  by the amount of fallback material  that it  is able to seize (Figure~\ref{fig:comparison}).
\end{itemize}

\begin{figure}[th!]
  \includegraphics[trim={0.8cm 1.2cm 0.8cm 1.4cm},clip,width=0.9\linewidth]{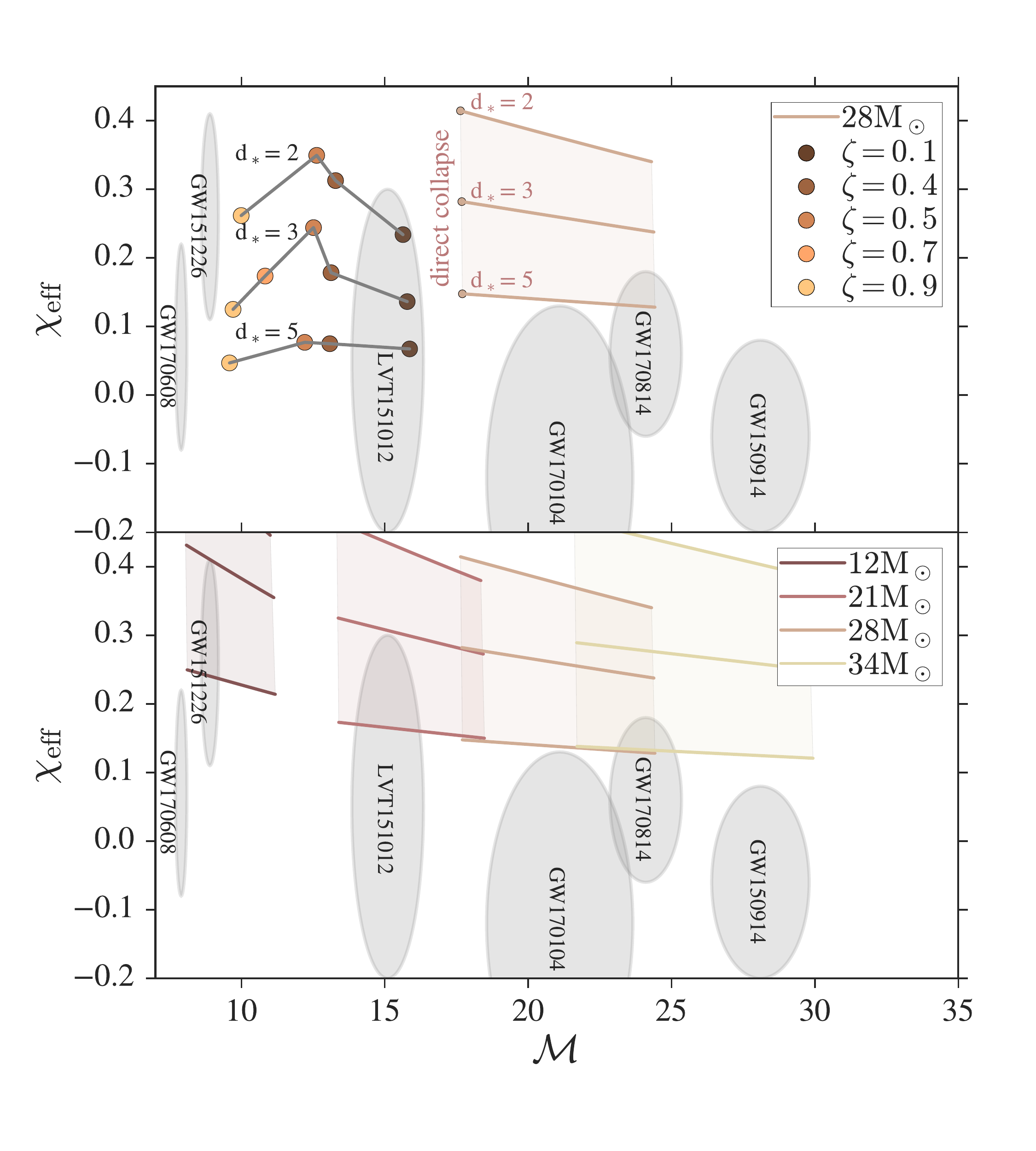} 
\caption{$\chi_{\rm eff}$ as a function of $\mathcal{M}$. {\it Top} panel: The results of our simulations ($28 M_\odot$  + $15 M_\odot$, $q=0.53$)  for varying $\zeta$ and $d$.  The shaded region show  the outcomes produced by  the direct collapse of the $28 M_\odot$ pre-SN helium star calculated  by varying $q$ from $ 0.53$ to $1$ and  $d$ from  $2R_\ast$ to $5R_\ast$. {\it Bottom} panel: The shaded quadrilateral regions show  systems produced by the direct collapse of  stars of varying $M_\ast=[12,34]M_\odot$, $q=[0.53,1]$ and $d=[2,5]R_\ast$. The corresponding stellar radius are  $0.66 R_\odot (12 M_\odot)$, $0.56 R_\odot  (21 M_\odot)$ and $0.49 R_\odot (34 M_\odot)$.
The shaded ellipses in both panels show the 90\% credibility intervals of the GW signals measured by LIGO.
 }
\label{fig:X_eff}
\end{figure}

In Figure \ref{fig:X_eff}  we present a comparison of our results ({\it upper} panel) in the context of both direct collapse solutions and current LIGO observations of binary BHs. Shown are $\chi_{\rm eff}$ as a function of the chirp mass, $\mathcal{M}$, of the resulting BH binary system. The shaded quadrilateral regions  ({\it upper} and {\it lower} panel) show  systems produced by the direct collapse of pre-SN helium stars of varying masses, whose structures have been taken from the  {\tt KEPLER} models of \citet{Woosley_2006}. The final spin of $M_2$ is calculated  using the radial  stellar profile and assuming rigid body rotation of the tidally synchronized SN progenitor. The pre-SN helium stars ($M_\ast$) are assumed to be  orbiting  around a BH  with $M_1=qM_\ast$ and $a_1$=0. The dependence of $\chi_{\rm eff}$ with $\mathcal{M}$ is obtained by varying $q$ from $ 0.53$ to $1$ in all cases, while the dependence of $\chi_{\rm eff}$ at a fixed  $\mathcal{M}$ is obtained by changing $d$ from $2R_\ast$ to $5R_\ast$ at constant  $q$. To facilitate comparisons, we plot as shaded ellipses the 90\% credibility intervals of the GW signals measured so far \citep{LIGO_GW151226,LIGO_GW150914,LIGO_GW170104,LIGO_GW170608,LIGO_GW170814}. 

Some points should be emphasized. The current LIGO observations are inconsistent with the direct collapse of pre-SN helium stars in close  binaries \citep{Kushnir_2016,Belczynski_2017b, Zaldarriaga_2018,Hotokezaka_2017a}. When the assumption of direct collapse is relaxed, the mass of $M_2$ can be altered  by small changes  on the explosion energy $\zeta$ while $a_2$ and $a_1$ (to a lesser extent) depend primarily on $d$ (Figure~\ref{fig:comparison}). For the specific $28 M_\odot$  + $15 M_\odot$ system studied here, we show that  changes in $\zeta$ alone can produce systems like LVT151012 ($\zeta=0.1$, $d_\ast=5$) or GW170608 ($\zeta=0.9$, $d_\ast=5$). For a fixed SN energy of $\zeta=0.9$, changes in the initial separation can, on the other hand, yield systems like GW170608 ($d_\ast=5$) or GW151226 ($d_\ast=2$).

Irrespective of the exact progenitor system, the processes discussed here implies  that the formation pathways 
of LIGO binary BHs are more complicated  than the  standard scenario suggests. But the effects are
especially interesting for weak SN explosions taking place  in close binary systems. Future LIGO observations 
can offer clues to the nature of the SN explosion leading to the formation of BHs, which is currently not well understood \citep{Perna2014,Sukhbold_2016,Raithel_2018}. For instance, 
GW170608 could be indicative of weak SN explosion of a more massive pre-SN progenitor system while 
GW151226  might arise due to direct collapse of a lighter, yet more compact progenitor system.  

The properties of LIGO sources in the ($\chi_{\rm eff}$,$\mathcal{M}$) plane  is diverse. One appealing aspect of the classical scenario  is that the great variety of binary and explosion parameters can probably help explain this diversity. Given the need for a large helium core mass in progenitors, BH formation may be favored not only by slow rotation but also by low metallicity \citep{Izzard_2004}.
Larger mass helium cores might have less energetic explosions but this is currently highly uncertain. Many massive stars may produce supernovae by forming neutron
stars in spherically symmetric explosions, but some may fail during neutrino energy deposition, forming BHs
in the centre of the star \citep{Fryer2012,Ugliano2012,Pejcha2015a} and possibly a wide range of weak SN explosions \citep{Moriya2010,Fryer2012, Dexter2013,Lovegrove2013,Batta_2017,Fernandez_2018}. One expects various outcomes ranging from very massive BHs with low spins (GW150914), to lighter and faster spinning  BHs (GW151226).  The number density of binary BHs of different masses and spins would provide a natural test to distinguish between different stellar explosion avenues.

\acknowledgments
We thank T. Fragos, W. Farr, C. Fryer and I. Mandel for useful conversations. 
We credit the Packard Foundation and the DNRF for support. Calculations were carried out at the UCSC's supercomputer Hyades (supported by NSF AST1229745)  and at the University of Copenhagen HPC centre (supported by  the Villum Fonden VKR023406). We are grateful to the Kavli Foundation and the DNRF for  sponsoring the 2017 Kavli Summer Program where part of this work was carried out.

\bibliography{SNspin_refs}

\end{document}